\documentstyle[amssymb,aps,preprint]{revtex}
\tightenlines

\begin{document}
\draft
\title{Localization of a random heteropolymer onto a surface}
\author{Semjon Stepanow and Alexander L. Chudnovskiy}
\address{Martin-Luther-Universit\"{a}t Halle-Wittenberg, Fachbereich Physik, D-06099\\
Halle, Germany}
\date{\today }
\maketitle

\begin{abstract}
We study the localization of a random heteropolymer onto an homogeneous
surface, the problem which is equivalent to the wetting of an interface at
disordered substrate in two dimensions, via replica trick by using the
Green's function technique. The exact treatment of one- and two-replica
binding states is used to compute the free energy of the random
heteropolymer. We present analytical results for two particular cases: {\it %
(i) }nearly statistically symmetric copolymer in the vicinity of the
threshold of the annealed problem, and {\it (ii)} the asymmetric polymer
with the interaction part of the annealed Hamiltonian being nearly zero. In
both cases the localization is due to two-replica binding states. In the
case {\it (i)} the two-replica binding state exists both above and below of
the one-replica binding state. In the case {\it (ii)} the energy of the
two-replica binding state at the transition is finite. A schematic phase
diagram of the localization-delocalization transition of the random
heteropolymer is suggested.
\end{abstract}


Adsorption of a polymer chain onto a surface is of large practical interest
ranging from biological physics to technological applications \cite
{gareletal89}-\cite{monthusetal00} (and citations therein). The adsorption
of a polymer with heterogeneous sequence structure is relevant in connection
with the study of the behavior of proteins near surfaces. The problem of
wetting in two dimensions \cite{fisher-me}-\cite{tangchate01} is closely
related to the adsorption of a symmetric heterogeneous polymer onto a
surface. The polymer adsorption has been studied in connection with
denaturation of double-stranded DNA in solution \cite{poland-scheraga70}- 
\cite{cule-hwa97} and growth problems \cite{kallabis-lassig95}.

The role of disorder on the absorption of the random heteropolymer remains a
subject of controversy. This question has been addressed recently by several
groups \cite{forgacsetal86,forgacsetal88,derridaetal92,tangchate01}. The
quenched part of disorder is concluded to be irrelevant in \cite
{forgacsetal86,forgacsetal88}, while on the contrary, the work \cite
{derridaetal92} shows that the quenched part of disorder shifts the
transition temperature of the localization. Very recently the similar
problem was considered in \cite{tangchate01} by using a functional
renormalization group. These authors predict that the localization
transition for the symmetric heteropolymer is of Kosterlitz-Thouless type.

In this Paper we study the adsorption of a random asymmetric heteropolymer
via replica trick by using the Green's function technique. In contrast to
the previous studies the Green's function method allows an exact
consideration of one-replica (1p) and two-replica (2p) binding states. Our
analysis shows that for statistically symmetric heteropolymer the 2p binding
state exists at the localization transition of the 1p binding state. The
exact solution of the two-replica problem at the point where the interaction
part of the annealed Hamiltonian is zero, indicates that the localization
transition is first-order. We incorporate the one-replica and two-replica
localized states into a novel and heuristic procedure to compute the
quenched free energy.

The partition function of a polymer containing $N$ segments interacting with
the surface is 
\begin{equation}
Z=\int Dz(s)\exp \left[ -\frac{1}{2l^{2}}\int_{0}^{N}ds\left( \frac{dz(s)}{ds%
}\right) ^{2}-\int_{0}^{N}dsV_{0}(z(s))-\int_{0}^{N}ds\zeta (s)V_{int}(z(s))%
\right] ,  \label{w1}
\end{equation}
where $V_{0}(z)=(\infty $, $z\leqslant 0$; $0,$ $z>0)$ is the repulsive
interaction potential with the wall, $l$ is the statistical segment length.
The heterogeneity of the polymer is described by random Gaussian variables $%
\zeta (s)$, which are characterized by the moments $\overline{\zeta (s)}%
=\zeta $, and $\overline{\zeta (s)\zeta (s^{\prime })}=\zeta ^{2}+\Delta
\delta (s-s^{\prime })$. If $\zeta =0$, then the number of the monomers
which are attracted or repelled from the well is on average the same, so
that the heteropolymer is statistically symmetric. On the contrary, if $%
\zeta \neq 0$, there is an excess of the monomers, which are repelled from ($%
\zeta >0$) or attracted to ($\zeta <0$) the well. In this case the
heteropolymer is asymmetric \cite{gareletal89}. The attractive interaction
with the surface will be modelled by the potential $V_{int}(z)=u\delta
(z-z^{0})$, where$\ z^{0}$ is small but nonzero (see below). This choice of
the potential allows the exact treatment of the problem with two replicas,
as will be shown below.

Imaging $s$ to be an axis perpendicular to $z$ we interpret Eq.(\ref{w1}) as
a partition function of a directed line interacting with a heterogeneous
substrate at $z=0$, which is the wetting problem. The random variables $%
\zeta (s)$ are now attributed to the substrate. The wetting interpretation
of the polymer problem is a particular case of the relation between the
polymer in $d$ dimensions and directed polymer in $d+1$ dimensions. In the
case of wetting the condition $\zeta \neq 0$ means that on average the
interface interacts with the substrate. If $\zeta >0$, then the interface is
repelled from the substrate.

Performing the average over $\zeta (s)$ by using the replica trick we obtain
the replica partition function as 
\begin{eqnarray}
Z_{n} &=&\overline{Z^{n}}=\int Dz_{a}(s)\exp
(-\sum\limits_{a=1}^{n}\int_{0}^{N}ds(\frac{1}{2l^{2}}(\frac{dz_{a}(s)}{ds}%
)^{2}+V_{0}(z_{a}(s))+\zeta V_{int}(z_{a}(s)))+  \nonumber \\
&&\frac{\Delta }{2}\sum\limits_{a,b=1}^{n}%
\int_{0}^{N}dsV_{int}(z_{a}(s))V_{int}(z_{b}(s))).  \label{w2}
\end{eqnarray}
The free energy has to be computed as follows 
\begin{equation}
-\beta F=\frac{\partial Z_{n}}{\partial n}\mid _{n=0}.  \label{w3}
\end{equation}

Due to the fact that (\ref{w2}) contains only the one-fold integral over $s$%
, the partition function $Z_{n}$ can be interpreted as the probability
amplitude of a quantum mechanical system of $n$ particles associated with
the Hamiltonian 
\begin{equation}
H_{n}=\sum\limits_{a=1}^{n}(-D\nabla _{a}^{2}+V_{0}(\ z_{a})-\beta \delta
(z_{a}-z^{0}))-\Delta u^{2}\sum\limits_{a<b}^{n}\delta (z_{a}-z^{0})\delta
(z_{b}-z^{0}),  \label{w5}
\end{equation}
where $D=l^{2}/2$, $\beta =\Delta u^{2}/2\delta _{0}-\overline{\zeta }$,
with $\delta _{0}=1/\delta (0)$ being of the order of magnitude equal to the
width of the potential well, and $\overline{\zeta }=\zeta u$. In the
following we will speak of particles instead of replicas. The first term in (%
\ref{w5}) is associated with the annealed average of the free energy. The
annealed part of $H_{n}$ is equivalent to the localization problem of a
quantum mechanical particle in the vicinity of the wall, and can be solved
exactly. The binding problem for (\ref{w5}) with $n=2$ can also be solved
exactly. The Hamiltonian (\ref{w5}) at given $n$ can be interpreted as a
Hamiltonian of $n$ polymers interacting with the surface. The 2nd term in
the first sum describes the attraction of the monomers to the surface, which
is independent of each other. The second sum in (\ref{w5}) gives an
additional attractive interaction, if the monomers belonging to different
polymers contact the surface simultaneously. We are not aware, if such an
interaction can be realized in reality. Eq.(\ref{w5}) at $n=2$ and for $%
\beta =0$ and $V_{0}(\ z_{a})=0$ is exactly the Hamiltonian of a quantum
mechanical particle in two dimensional delta potential. As it is well-known
from text books the binding state in this case exists for infinitesimally
weak potential. While $\sum_{a=1}^{n}V_{0}(\ z_{a})$ does not possess the
radial symmetry, the Hamiltonian (\ref{w5}) at $n=2$ and $V_{0}(\ z_{a})\neq
0$ does not correspond to a quantum mechanical problem in a radial symmetric
two dimensional potential well with an impenetrable core at the origin. The
Hamiltonian $H_{n}$ at $n=2$, $V_{0}(\ z_{a})=0$ and $\beta =0$ is related
to the Poland-Sheraga model \cite{poland-scheraga70}. The essential
difference to the Poland-Sheraga model consists in the fact that all
contacts occur at $z=z_{0}$, i.e. the model (\ref{w5}) neglects the wiggling
of the zipped polymer pair. It is evident that the last term in (\ref{w5})
favors the localization of replicas.

To this end it is convenient to consider the one-replica Green's function, $%
G(z_{1},N;z_{1}^{0})\equiv \langle \delta (z(N)-z_{1})\delta
(z(0)-z_{1}^{0})\rangle $, associated with the annealed part of the
Hamiltonian (\ref{w5}). The Laplace transform with respect to $N$ (the
variable $N$ plays the role of the imaginary time for the quantum mechanical
particle) of the perturbation expansion of $G(z_{1},N;z_{1}^{0})$ in powers
of the attraction strength $\beta $ is a geometric series, which is summed
as 
\begin{equation}
G(z_{1},p;z_{1}^{0})=G_{0}(z_{1},p;z_{1}^{0})+\beta \frac{G_{0}(\
z_{1},p;z_{0})G_{0}(z_{0},p;z_{1}^{0})}{1-\beta G_{0}(z_{0},p;z_{0})},
\label{w6}
\end{equation}
where $G_{0}(z_{1},p;z_{1}^{0})=(\exp (-\left| z_{1}-z_{1}^{0}\right| \sqrt{%
p/D})$ $-$ $\exp (-\left| z_{1}+z_{1}^{0}\right| \sqrt{p/D}))$ $/\sqrt{4Dp}$
is the Laplace transform of the Green's function of the diffusion equation
in the half space ($z\geqslant 0$) with the Dirichlet boundary condition at $%
z=0$. The equation 
\begin{equation}
1-\beta G_{0}(z_{0},p;z_{0})=1-\beta (4Dp)^{-1/2}(1-\exp (-2z_{0}\sqrt{p/D}%
))=0,  \label{w6-1}
\end{equation}
which is the denominator of the 2nd term in the r.h.s. of Eq.(\ref{w6}), is
the energy eigenvalue condition for 1p (one-replica or one-particle)\
localized state. Identifying $D$ as $\hbar ^{2}/2m$, Eq.(\ref{w6-1})
coincides exactly with the eigenenergy condition for the localization of a
quantum mechanical particle in an attractive Delta-potential placed at the
distance $z_{0}$ from the wall. The localized state corresponds to the
solution of (\ref{w6-1}) $p_{c}>0$. The energy of the localized state is
given by $E_{1,0}=-p_{c}$. It is easy to see from (\ref{w6-1}) that the
localized state exists for $\beta >\beta _{c}=D/z_{0}$. The inverse Laplace
transform of $G(z_{0},p;z_{0})$ for weak binding ($2z_{0}\sqrt{p/D}\leqslant
1$) and $\beta >\beta _{c}$ is obtained from (\ref{w6}) as

\begin{eqnarray}
G(z_{0},N;z_{0}) &\simeq &\frac{\sqrt{D}}{\beta z_{0}}(\sqrt{p_{c}}\exp
(Np_{c})\,+\frac{1}{\sqrt{\pi N}\,}-\frac{\,p_{c}\,z_{0}}{\sqrt{D}}\exp
(Np_{c})-\sqrt{\frac{p_{c}}{\pi DN}}z_{0}+  \nonumber \\
&&\,\sqrt{p_{c}}\,\exp (Np_{c})%
%TCIMACRO{\func{erf}}%
%BeginExpansion
\mathop{\rm erf}%
%EndExpansion
(\sqrt{N\,p_{c}})-\frac{\,p_{c}\,z_{0}\,}{\sqrt{D}}\exp (Np_{c})%
%TCIMACRO{\func{erf}}%
%BeginExpansion
\mathop{\rm erf}%
%EndExpansion
(\sqrt{N\,p_{c}})),  \label{w7}
\end{eqnarray}
where $p_{c}=(\beta /\beta _{c}-1)^{2}D^{3}/(\beta ^{2}z_{0}^{4})$. In
computing (\ref{w7}) we neglected the term proportional to $\delta (N)$ i.e.
we consider $N$ in (\ref{w7}) to be positive. Taking into account the delta
function in computing the two-replica binding state in Eqs.(\ref{w9}-\ref
{w10}) leads to non significant changes.

To study the effect of the non-diagonal part of $H_{n}$ in the case $n=2$ we
will consider the connected part of the two-replica Green's function $%
G_{2,c}(z_{1},z_{2},N;z_{1}^{0},z_{2}^{0},0)\equiv \langle \delta
(z_{1}-z_{a}(N))\delta (z_{2}-z_{b}(N))\delta (z_{1}^{0}-z_{a}(0))\delta
(z_{2}^{0}-z_{2}(0))\rangle _{c}$, where $a$ and $b$ denote the replica
indices $(a\neq b)$. The perturbation expansion of the Green's function $%
G_{2,c}(z_{1},z_{2},N;z_{1}^{0},z_{2}^{0},0)$ in powers of the interaction
(third and fourth terms in (\ref{w5})) is represented graphically in Fig.1.
The dotted lines are associated with the two-replica interaction given in
Eq.(\ref{w5}). The ends of the dotted lines are associated with $z_{0}$ and
the arc length $s_{i}$, which are ordered from the left to the right. An
integration over $s_{i}$ has to be performed. Each part of the continuous
line between two consecutive dotted lines is associated with the one-replica
Green's function $G(z^{0},s_{i};z^{0},s_{i-1})$. The left (right) external
lines are associated with $G(z,N;z^{0},s_{i})$ ($G(z_{0},s_{1};z^{0},0)$) ($%
z $ be $z_{1}$, or $z_{2}$, while $z^{0}$ be $z_{1}^{0}$, or $z_{2}^{0}$).
The graphical expansion in Fig.1, which visualizes the effect of two-replica
interaction in $H_{2}$ in terms of space-time ($N=-it$) trajectories, shows
that both trajectories contact the surface at the same time. Thus, the
return probability to have two consecutive contacts is the square of that
for one particle. This suggests that the localization of two particles
interacting according to (\ref{w5}) is closely related to the localization
of one particle in two dimensions. The integral associated with a graph in
Fig.1 is a folding, so that the Laplace transform with respect to $N$
reduces the perturbation expansion in Fig.1 to a geometrical series, which
is summed as 
\begin{equation}
G_{2,c}(z_{1},z_{2},p;z_{1}^{0},z_{2}^{0})=\alpha \frac{\widetilde{G}_{2}(\
z_{1},z_{2},p;z_{0},z_{0})\widetilde{G}%
_{2}(z_{0},z_{0},p;z_{1}^{0},z_{2}^{0})}{1-\alpha \widetilde{G}%
_{2}(z_{0},z_{0},p;z_{0},z_{0})},  \label{w8}
\end{equation}
where $\alpha =\Delta u^{2}$, and $\widetilde{G}%
_{2}(z_{1},z^{2},p;z_{1}^{0},z_{2}^{0})$ is the Laplace transform of the
product of two one-replica Green's functions $%
G(z_{1},N;z_{1}^{0})G(z_{2},N;z_{2}^{0})$. The denominator on the r.h.s. of
Eq.(\ref{w8}) gives the eigenvalue condition for the two-replica (particle)
bound state. We did not succeed to analyze the latter analytically, so that
now we will consider the following particular cases: {\it (i)} approximately
statistically symmetric polymer in the vicinity of the localization
threshold of the annealed problem, and {\it (ii)} the asymmetric polymer
under condition that the interaction part of the annealed Hamiltonian is
approximately zero. The latter case can be realized by tuning the asymmetry
parameter $\zeta $. The two-replica partition function for directed polymers
with random interactions was previously studied in \cite{bhattacharjee93}- 
\cite{mukherji93}.

To ensure the existence of Laplace transform of the one-replica Green's
function squared, we introduce a short-time cutoff by replacing $1/N$ by $%
1/(N+a)$. The cutoff $a$ along the polymer can be eliminated in favor of the
transversal length $a_{0}$ via $a=a_{0}^{2}/4D$\footnote{%
Notice that the case, when only the first term in (\ref{w11}) is present,
corresponds to localization of a QM particle in a shallow 2d potential well.
The above procedure gives an exact solution of the problem, if one
identifies the length $a_{0}$ with the width of the potential well.}. The
eigenvalue condition for the two-replica localized state, which consists in
equality of the denominator of Eq.(\ref{w8}) to zero, is obtained for small $%
p_{c}$, i.e. in the vicinity of the localization transition of the annealed
problem, as 
\begin{eqnarray}
0 &=&1-\frac{D\,\,\alpha \,}{\pi \,z{_{0}}^{2}\,\beta ^{2}}\exp
(a_{0}^{2}\,p/4\,D)\Gamma (0,a_{0}^{2}\,p/4\,D)+  \nonumber \\
&&\frac{p_{c}^{2}\,D\,\alpha }{\pi \,z{_{0}}^{2}\,\beta ^{2}}\left( \frac{%
-2\,\pi }{\sqrt{p}}+\frac{2z_{0}}{\sqrt{D}}\exp (a_{0}^{2}\,p/4\,D)\Gamma
(0,a_{0}^{2}\,p/4\,D)\,\,\right) \,,  \label{w9}
\end{eqnarray}
where $\Gamma (0,x)$ is the incomplete gamma function. Eq.(\ref{w9}) yields
for $p$ in the vicinity of the one-replica binding transition, i.e. for
small $p_{c}$ 
\begin{equation}
p_{2,c}=\frac{4\,D\exp (-\gamma )}{a_{0}^{2}}\exp (-\frac{\pi \,z_{0}^{2}\,{%
\beta }^{2}}{D\,\alpha })\,,  \label{w10}
\end{equation}
where $\gamma $ is the Euler number. The energy $E_{2,c}$ of the two-replica
bound state is $-p_{2,c}$. For the symmetric case, $\alpha =2\beta \delta
_{0}$, $p_{2,c}$ decreases with decreasing $\delta _{0}$ at fixed $\beta
\sim \beta _{c}$ and $z_{0}$. This ensures the validity of the condition $%
2z_{0}\sqrt{p/D}\leqslant 1$ (weak binding) that we used to derive Eq.(\ref
{w10}). A similar analysis slightly above the annealed threshold also
results in Eq.(\ref{w10}), so that the two-replica bound state exists both
below and above the threshold of the one-replica bound state. The result (%
\ref{w10}) shows that the two-replica bound state already exists at the
one-replica localization transition. This is very reasonable and can be
explained qualitatively as follows. The individual interactions with the
surface contained in the annealed part of the Hamiltonian (\ref{w5}) result
in an increase of the probability to find the monomers of the polymer pair
in the vicinity of the surface. This compensates the decrease of the
probability, which is due to the wall potential $\sum_{a=1}^{n}V_{0}(\
z_{a}) $, and thus shifts the threshold to the lower values.

The asymmetric case {\it (ii), }$\beta \rightarrow 0${\it ,} where no
one-replica localized states exist can be realized by tuning the asymmetry
parameter $\zeta $ as it is seen from the definition of $\beta $. For small $%
\beta $ the 1p Green's function may be approximated by its bare value, so
that we obtain 
\begin{equation}
G^{2}(z_{0},N;z_{0})=\frac{1}{4\pi DN}(1-2\exp (-\frac{z_{0}^{2}}{2DN})+\exp
(-\frac{2z_{0}^{2}}{2DN})).  \label{w11}
\end{equation}
As above we replace $1/N$ in the first term on the right-hand side of (\ref
{w11}) by $1/(N+a)$, where the cutoff $a$ along the polymer can be
eliminated in favor of the length $a_{0}$ via $a=a_{0}^{2}/4D$. The $z_{0}$%
-dependent terms in (\ref{w11}) are due to the boundary condition at the
surface $z=0$. In the case of adsorption onto an interface only the term $%
1/(4\pi DN)$ will appear in Eq.(\ref{w11}), so that in this case the
two-replica bound state will exactly coincide with that in a shallow
potential well in two dimensions. Using (\ref{w11}) we obtain from (\ref{w8}%
) the eigenvalue condition for the two-replica bound state as 
\begin{equation}
1+4\alpha _{1}K_{0}(4\sqrt{\widetilde{p}/\sigma ^{2}})-2\alpha _{1}K_{0}(4%
\sqrt{2\widetilde{p}/\sigma ^{2}})-\alpha _{1}\exp (\widetilde{p})\Gamma (0,%
\widetilde{p})=0,  \label{w12}
\end{equation}
where $K_{0}(x)$ is the modified Bessel function of the second kind, $%
\widetilde{p}=pa_{0}^{2}/D$, $\alpha _{1}=\alpha /4\pi D$, $\sigma
=a_{0}/z_{0}$. Identifying the cutoff $a_{0}$ with $z_{0}$ gives $\sigma =1$%
. It appears that the results are not sensitive to the choice of $a_{0}$.
The numerical analysis of Eq.(\ref{w12}) for different $\sigma $ ($=0.8$, $1$%
, $1.2$) yields the critical value $\alpha _{1}^{c}$ for the localization
transition ($=0.5825$, $0.7857$, $1.085$). It appears that at the transition 
$\widetilde{p}_{2,c}^{0}$ has the finite value ($=0.0001$, $0.00195$, $%
0.01963$), i.e. the binding transition is a first-order transition. Above
the transition, $\alpha _{1}>\alpha _{1}^{c}$, there are two solutions for $%
\widetilde{p}_{2,c}$: $\widetilde{p}_{2,c}<\widetilde{p}_{2,c}^{0}$ and $%
\widetilde{p}_{2,c}>\widetilde{p}_{2,c}^{0}$. According to the ground state
dominance argument the larger value governs the behavior of the polymer for
large $N$. The approximative consideration based on the Taylor expansion of
the eigenvalue condition (\ref{w12}) for small $\widetilde{p}$%
\begin{equation}
1-\alpha _{1}(\gamma +\ln 2-2\ln \sigma )+\alpha _{1}\widetilde{p}(-1+\gamma
+8\ln 2/\sigma ^{2}+\ln \widetilde{p})+...=0  \label{w12a}
\end{equation}
is in agreement with the results of numerical consideration of Eq.(\ref{w12}%
). The reason of the unusual first-order transition is due to the term $%
\widetilde{p}\ln \widetilde{p}$ in Eq.(\ref{w12a}). The latter is
responsible for that the l.h.s. of (\ref{w12a}) has a minimum at finite $%
\widetilde{p}$, which leads to the first-order transition. The physical
reason of the first-order transition is due to the boundary condition at the
wall $z=0$, which results in a reduction of number of conformations, and
drives the transition to be first-order. As we stressed above the problem
associated with the Hamiltonian $H_{n}$ given by Eq.(\ref{w5}) for number of
replicas $n=2$ and $\beta =0$ is closely related to the Poland-Sheraga model 
\cite{poland-scheraga70}, which was recently studied in \cite
{causo/coluzzi/grassberger}-\cite{garel/monthus/orland01} by taking into
account the excluded-volume interaction between denaturated loops and the
rest of the chain. The excluded volume interaction reduces the number of
conformations and drives the transition to the first order. This is similar
to the two-replica localized state, where the reduction of the number of
conformations is due to the effect of the wall potential $V_{0}(z)$.

We now will consider the computation of the free energy by using the replica
formula (\ref{w3}) under taking into account the one-replica and two-replica
binding states. In the case, if only 1p (one-replica) binding states exist,
Eq.(\ref{w3}) gives straightforwardly $-\beta F=p_{c}$. However, the
situation is nontrivial when a 2p (two-replica) binding state exists. Taking
into account the two-replica states in the two-pair approximation we obtain
the partition function $Z_{n}$ as 
\begin{equation}
Z_{n}=Z_{1}^{n}+\frac{n(n-1)}{2}Z_{2,c}Z_{1}^{n-2}+...=\exp (np_{c}N)+\frac{%
n(n-1)}{2}\exp (p_{2,c}N+(n-2)p_{c}N)+...,  \label{w13}
\end{equation}
where we have taken into account that $Z_{1}$ and $Z_{2,c}$ behave for large 
$N$ as $\exp (p_{c}N)$ and $\exp (p_{2,c}N)$, respectively, when both 1p and
2p bound states exist. If no 1p bound state exist, then $Z_{1}=1$. The next
terms in (\ref{w13}) contain contribution of ternary and higher pairs. Using
(\ref{w3}) and (\ref{w13}), the free energy is obtained to be proportional
to $\exp ((p_{2,c}-2p_{c})N)$, hence the free energy is not extensive
quantity for $p_{2,c}-2p_{c}>0$ . This shows that the two-pair approximation
(\ref{w13}) is insufficient for computing the free energy. The problem is
due to the fact that the exponentially increasing term in (\ref{w13}) that
originates from the two-particle bound state does not contain the factor $n$%
, as it is the case for the 1p bound state. The factor $n$ in the
exponential of the latter ensures that it disappears in the limit $%
n\rightarrow 0$. To overcome the difficulty we suggest to take into account
in the expansion of $Z_{n}$ the term containing the maximal number of
unconnected pairs. Then, instead of (\ref{w13}) we obtain 
\begin{equation}
Z_{2n}=Z_{1}^{2n}+2^{-n}\frac{\Gamma (2n+1)}{\Gamma (n+1)}Z_{2,c}^{2n/2}+...,
\label{w14}
\end{equation}
where we consider even $n$. The factor $2^{-n}\Gamma (2n+1)/\Gamma (n+1)$ in
Eq.(\ref{w14}) is the analytical continuation of $(2n-1)!!$ for arbitrary $n$%
. The free energy is now obtained from Eq.(\ref{w3}) as 
\begin{equation}
\frac{-\beta F}{N}=p_{c}+\frac{1}{2}p_{2,c}+...  \label{w15}
\end{equation}
The localization length $\xi _{loc}$ can be computed by using $p_{2,c}$ as $%
\xi _{loc}\simeq (p_{2,c}/D)^{-1/2}$. The regime {\it (ii)} considered above
is obtained from (\ref{w15}), if one puts $p_{c}=0$. Notice the plus sign in
front of the 2p energy in (\ref{w15}). The computation of the 2p energy in
the two-pair approximation (\ref{w13}) gives in the limit $n\rightarrow 0$
instead of (\ref{w15}) the minus sign in front of the second term. The
inconsistency of this result is clearly demonstrated by the non
extensiveness of the free energy computed by using Eq.(\ref{w3}). The
procedure of taking into account the terms with maximal number of pairs can
be justified by the following ground state dominance like argument: the
term, $\exp (np_{2,c}N)$, dominates over $\exp (p_{2,c}N)$ for large $N$ and 
$n\geqslant 1$. The condition $n\geqslant 1$ is demanded in the procedure of
introduction of the replica trick by considering the partition function $%
Z_{n}$ with$\ n$ being positive integer. For the peculiarities of the limits 
$N\rightarrow $ $\infty $ and $n\rightarrow 0$ in the replica treatment of
the directed polymer in disordered media see \cite{kardar-leshouches}. The
extensiveness of the free energy is a posteriori justification of the above
heuristic procedure.

We now will use the results of the study of 1p and 2p bound states of the
replica Hamiltonian (\ref{w5}) to construct the phase diagram of the
localization of the heteropolymer in variables $\overline{\zeta }$ and $%
\alpha _{1}=\Delta u^{2}/4\pi D$, which is shown schematically in Fig.2. We
know the behavior at $\alpha _{1}^{c}$ and $\alpha _{1}^{a}$ from the study
of the cases {\it (ii)} and {\it (i)}. The latter corresponds to the
localization transition of the annealed Hamiltonian (1p bound state) in the
symmetric case ($\overline{\zeta }=0$). The dotted straight line is the
localization line of the annealed Hamiltonian; $\overline{\zeta }_{c}$ $%
=-D/z_{0}$ is the value of the asymmetry parameter in the limit $\alpha
_{1}\rightarrow 0$. Since the random heteropolymer can arrange at the
surface in such a way that pieces of the polymer which are attracted to the
surface are in contact with the latter, while the pieces which are repelled
are in the loops and tails (see \cite{joanny94} for a related discussion),
the random quenched heteropolymer is expected to localize easier than the
homopolymer or even\ easier than the annealed heteropolymer. Due to this the
quenched part of the Hamiltonian (\ref{w5}), which is responsible for 2p
bound states, shifts the critical temperature to higher values, so that the
localization line will cut the horizontal axis on the left from the point $%
\alpha _{1}^{a}$. The continuous line cannot end in the dotted line, while
the analysis carried out in {\it (i)} in the vicinity of $\alpha _{1}^{a}$
applies also along the dotted line, so that the 2p localized state exists
there. Thus, the continuous line will be on the left of the dotted line, and
it will end in $\overline{\zeta }_{c}$. In order to make our prediction that
the localization transition at $\alpha _{1}^{c}$ is first-order compatible
with the prediction in \cite{tangchate01} that the transition at $\overline{%
\zeta }=0$ is of Kosterlitz-Thouless type, one should assume that on the
phase boundary in Fig.2 there is a tricritical point separating the KT point
at $\overline{\zeta }=0$ with the first-order transition at $\alpha _{1}^{c}$%
.

In summary, we have considered adsorption of a random heteropolymer onto a
homogeneous surface via replica trick by using the Green's function
technique. We use the exact treatment of one- and two-replica binding states
of the replica Hamiltonian to compute the free energy of the random
heteropolymer. We have considered analytically two particular cases: {\it (i)%
} almost statistically symmetric polymer in the vicinity of the threshold of
the annealed problem, and {\it (ii)} the asymmetric polymer where the
interaction part of the annealed Hamiltonian is nearly zero. In the former
case we have obtained that the localization of the polymer is due to the 2p
binding state, which exists both above and below of the 1p binding state,
which corresponds to the localization transition of the annealed problem. We
have obtained that in the case {\it (ii)} the energy of the 2p binding state
at the transition is finite, i.e. the localization is a first-order
transition. A schematic phase diagram of the localization-delocalization
transition of the random heteropolymer is suggested.

\acknowledgments
We acknowledge a support from the Deutsche Forschungsgemeinschaft, SFB 418
and \ grant Ste 981/1-1. S.S. acknowledges stimulating discussions with T.
Garel and H. Orland, especially on the relation between polymer adsorption
problem and the wetting problem. We thank R. Netz for bringing the reference 
\cite{kafri-mukamel-peliti00} to our attention.

\newpage Figure captions

Fig.1 The perturbation expansion of the connected part of the two-replica
Green's function.

Fig.2 The phase diagram of the localization-delocalization transition.

\end{document}